\documentclass[sigconf,natbib=false]{acmart}
\usepackage[style=ACM-Reference-Format,backend=bibtex,sorting=none]{biblatex}

\usepackage{multirow}
\usepackage{breqn}
\AtBeginDocument{%
  \providecommand\BibTeX{{%
    \normalfont B\kern-0.5em{\scshape i\kern-0.25em b}\kern-0.8em\TeX}}}

\setcopyright{none}
\renewcommand\footnotetextcopyrightpermission[1]{}

\begin{document}
\title{MRAM-based Analog Sigmoid Function for \\In-memory Computing}

\author{Md Hasibul Amin, Mohammed Elbtity, Mohammadreza Mohammadi, Ramtin Zand}
\affiliation{%
  \institution{Department of Computer Science and Engineering, University of South Carolina, Columbia, SC, 29201
  \country{USA}
  }
}


\begin{abstract}

We propose an analog implementation of the transcendental activation function leveraging two spin-orbit torque magnetoresistive random-access memory (SOT-MRAM) devices and a CMOS inverter. The proposed analog neuron circuit consumes $1.8-27\times$ less power, and occupies $2.5-4931\times$ smaller area, compared to the state-of-the-art analog and digital implementations. Moreover, the developed neuron can be readily integrated with memristive crossbars without requiring any intermediate signal conversion units. The architecture-level analyses show that a fully-analog in-memory computing (IMC) circuit that use our SOT-MRAM neuron along with an SOT-MRAM based crossbar can achieve more than 1.1$\times$, 12$\times$, and 13.3$\times$ reduction in power, latency, and energy, respectively, compared to a mixed-signal implementation with analog memristive crossbars and digital neurons. Finally, through cross-layer analyses, we provide a guide on how varying the device-level parameters in our neuron can affect the accuracy of multilayer perceptron (MLP) for MNIST classification.

\end{abstract}





\maketitle
\pagestyle{plain}

\section{Introduction}
In-memory computing (IMC) systems, as an alternative for von Neuman architectures, aim to eliminate the expensive data movement between processor and memory in data-intensive applications such as machine learning (ML) by implementing computation where the data exist. IMC architectures leverage massive parallelism in memristive crossbars and analog computation to aggressively reduce the computational time complexity of matrix-vector multiplication (MVM) operations in ML workloads. However, there are still a wide range of computational tasks in ML workloads that cannot be performed in crossbars, e.g., activation functions and subsampling layers. The solution offered in most of the existing IMC architectures such as ISAAC \cite{ISAAC}, PUMA\cite{PUMA}, and PRIME \cite{chi2016prime}, is to design digital circuits for some of these functions, and place them near crossbars. Although this approach reduces the off-chip data transfer overhead between processor and memory for commonly-used functions, it requires analog-to-digital converter (ADC) and digital-to-analog converter (DAC) units to transfer the data between crossbars and digital functional units. According to a research by HP Labs \cite{DPE}, the signal converters and peripheral circuitry in IMC circuits can contribute to up to $\>$90\% and $\>$95\% of the total power consumption and area occupation, respectively. Therefore, reducing the need for signal conversion units in an IMC architecture can potentially lead to significant energy improvements.

Transcendental activation functions such as sigmoid and hyperbolic tangent (tanh) provide the desired nonlinear behavior which can be useful for a variety of ML models such as multilayer perceptron (MLP) and Long Short-Term Memory (LSTM). However, due to their hardware implementation challenges, alternative nonlinear activation functions such as rectified linear unit (ReLU), first proposed in \cite{AlexNet}, have become popular in modern deep learning models such as convolutional neural networks (CNNs) \cite{VGG-16, szegedy2015going}. Hardware implementation of sigmoid and tanh functions have been widely investigated in the digital domain.

Direct implementations based on high-order polynomials, Tylor's series, and Maclaurin series offer high precision at the expense of high power consumption \cite{1292370}. On the other hand, CORDIC-based hardware implementation \cite{9180864} can achieve high accuracy and reduced area, however they have a high latency. In the PWL technique the non-linear function is separated into several linear segments to reduce implementation complexity \cite{9218549}. Higher precision can be accomplished by increasing the number of linear segments at the cost of more hardware resource utilization. Among the several function approximation approaches, LUT methods are the fastest \cite{5642617}, but it requires large amount of memory to provide high accuracy.

Although digital hardware can provide an exact implementation of the ideal sigmoid function and is more robust to noise, its high power, area, and latency overheads have motivated the research on analog transcendental activation functions. In \cite{neuron1}, three NMOS and three PMOS transistors are carefully sized to approximate a sigmoid activation function with only one reference voltage to bias the transistors. In \cite{neuron2}, an NMOS and a PMOS transistors with large width/length ratios are utilized along with a resistor load to realize $tanh$-like activation function. While, both of these design achieve significant reductions in power and area compared to digital implementations, they still require large transistors. In this paper, we leverage spin-orbit torque magnetoresistive random-access memory (SOT-MRAM) devices and a simple CMOS inverter to develop a sigmoid activation function that is compatible with memristive crossbars. We exhibit the direct and indirect energy and performance benefits of our proposed analog neuron for the IMC architectures through the circuit implementation of a fully-analog MLP circuit and comparing it with existing mixed-signal designs. While MRAM technology has been previously used for realizing probabilistic sigmoidal activation functions in binary stochastic neurons \cite{zand2018composable, ZandGLSVLSI}, this is the first work which leverages MRAM devices to implement a transcendental activation function.

\section{SOT-MRAM Technology}

\begin{figure}
\centering
\includegraphics[width=3.4in]{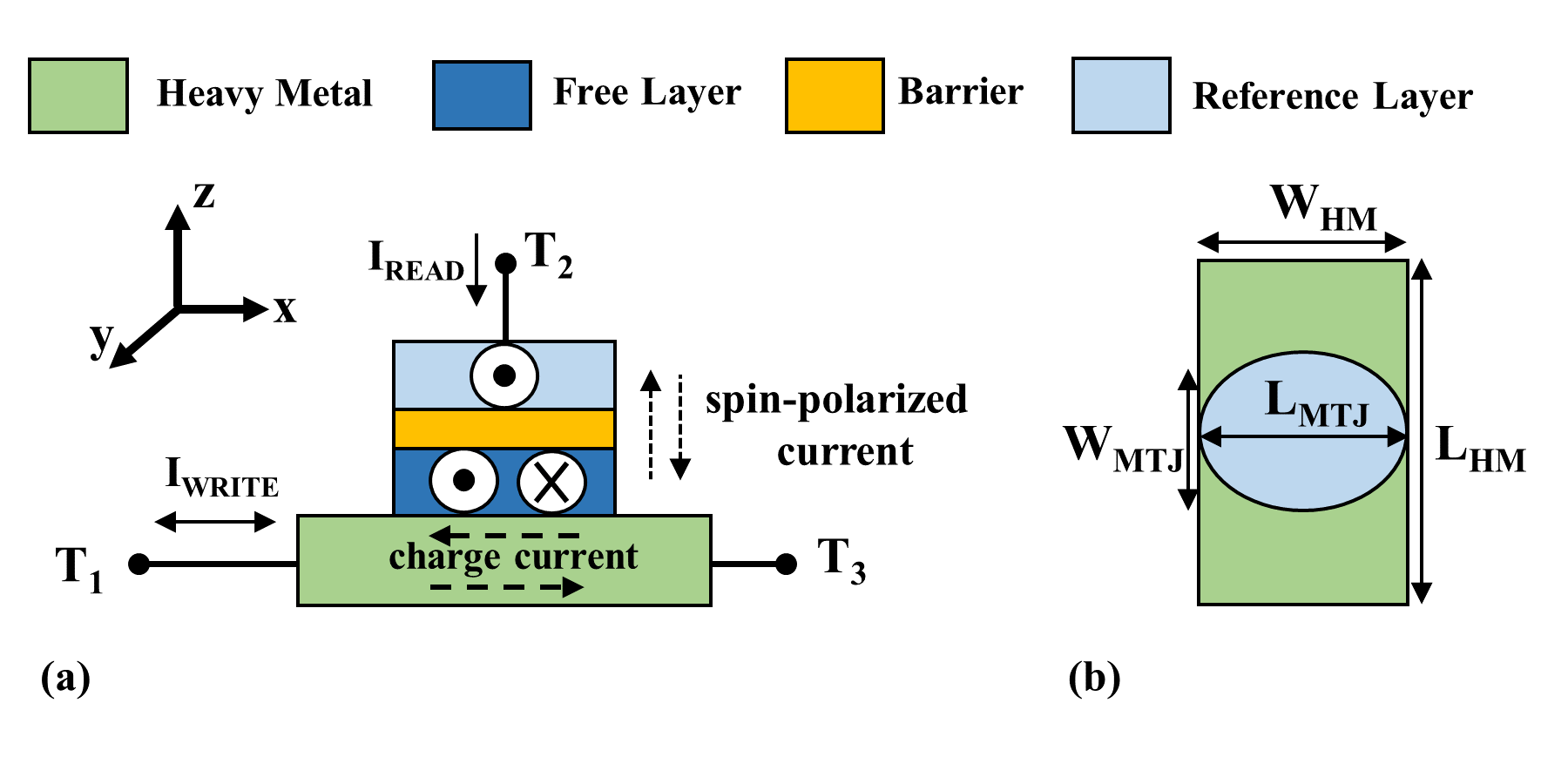}
\caption{(a) SOT-MRAM cell. A charge current along $+x/-x$ generates a spin-polarized current in $+z$ direction producing a spin torque that switches the magnetic direction of the free layer in $+y/-y$ direction. (b) SOT-MRAM Top view.}
\label{fig:sotmram}
\end{figure}

MRAM device is one of the promising memristive technologies that have been widely investigate for the IMC circuits and architectures \cite{ZandGLSVLSI,IMAC}. The magnetic tunnel junctions (MTJs) are the main building block in MRAM devices. MTJs include two ferro-magnetic (FM) layers that are separated by a thin oxide layer as shown in Figure \ref{fig:sotmram}. The magnetization direction of electrons in one of the FM layers (reference layer) is fixed, whereas the direction of electrons in the other FM layer (free layer) can be switched. The resistance levels of MTJ are determined by the angle ($\theta$) between the magnetization orientation of the FM layers. The following equations can be used to calculate the resistance of the MTJ in parallel (P) and antiparallel (AP) magnetization configurations \cite{Zhang2012CompactModeling, zand2018fundamentals}:

\begin{equation} 
\label{EqR} 
\small 
\begin{aligned}
R(\theta) = & \frac{2R_{MTJ}(1 + TMR)}{2 + TMR ( 1 + \cos\theta)}= \begin{cases} 
R_P=R_{MTJ}, & \theta=0  \\ 
R_{AP}=R_{MTJ}(1+TMR), & \theta=\pi 
\end{cases} 
\end{aligned}
\end{equation}

\begin{equation} \small \label{EqTMR} TMR(T,V_b)= \frac{TMR_0/100}{1+(\frac{V_b}{V_0})^2}\    \end{equation}

\noindent where $R_{MTJ} = \frac{RA}{Area}$. The MTJ's resistance-area product (RA) value is determined by the material composition of its layers. TMR stands for tunneling magnetoresistance, and it is determined by temperature (T) and bias voltage ($V_b$). $TMR_0$ is a material-dependent constant, and $V_0$ is a fitting parameter.

\begin{table}
\centering
\small
\caption{Parameters of the SOT-MRAM device \cite{zand2018fundamentals}}
\label{table:sheparameters}
\begin{tabular}{c c c} \hline  
{\bf Parameter} & {\bf Description} & {\bf Value} \\ \hline
\multirow{2}{*}{$MTJ_{Area}$} &  \multirow{2}{*}{$l_{MTJ}\times w_{MTJ} \times \frac{\pi}{4}$} & \multirow{2}{*}{$50nm \times 30nm \times \frac{\pi}{4}$}  \\ 
		{}&{}&{} \\ 
\multirow{2}{*}{$HM_{V}$} &  \multirow{2}{*}{$l_{HM}\times w_{HM} \times t_{HM}$} & \multirow{2}{*}{$100nm \times 50nm \times 3nm$}  \\ 
{}&{}&{} \\ 
$\rho_{HM}$ &    heavy metal resistivity & $200\mu\Omega.cm$  \\
{$RA$}&    resistance-area product & 10 $\Omega.\mu m^2$  \\
{$V_{0}$}&    Fitting parameter & 0.65  \\ 
{$TMR_0$}&    tunneling magnetoresistance & {100}  \\ \hline
\end{tabular}
\end{table}

SOT-MRAM device \cite{Liu2012} is a class of MRAM technology, in which the magnetization direction of the free layer can be changed by means of a spin-polarized current produced by a charge current passing through a heavy metal (HM), as shown in Fig. \ref{fig:sotmram}. The resistance of the HM can be calculated using the below equation:

\begin{equation} \small \label{EqRhm} R_{HM}=\rho_{HM}.l_{HM}/w_{HM}.t_{HM}    \end{equation}

\noindent where $\rho_{HM}$, $l_{HM}$, $w_{HM}$, $t_{HM}$ are the resistivity, length, width, and thickness of the HM, respectively. Here, we use the aforementioned MTJ and HM equations to create a Verilog-A model of the SOT-MRAM device for circuit simulations using the parameters mentioned in Table \ref{table:sheparameters} as default values.

\section{MRAM-based Sigmoid Neuron}

\begin{table*}[]
\caption{Drain-to-source currents for different operating regions of MP and MN.}
\vspace{-2mm}
\label{tab:equation}
\centering
\begin{tabular}{lcccccc}
\hline
& Cut-off && Triode && Saturation \\\hline
$I_{DSP}$ & $0$ && $\beta_P[(V_{INV}-V_{DD}-V_{TP})(V_{OUT}-V_{DD})- \frac{(V_{OUT}-V_{DD})\textsuperscript2}{2}]$ && $-\frac{\beta_P}{2}(V_{INV}-V_{DD}-V_{TP})\textsuperscript2$ \\
$I_{DSN}$ & $0$ && $\beta_N[(V_{INV}-V_{SS}-V_{TN})(V_{OUT}-V_{SS})- \frac{(V_{OUT}-V_{SS})\textsuperscript2}{2}]$ && $-\frac{\beta_N}{2}(V_{INV}-V_{SS}-V_{TN})\textsuperscript2$ \\
\hline
\end{tabular}
\end{table*}

\begin{table*}[]
\caption{Conditions for different operating regions of MP and MN with $V_{TP}$ and $V_{TN}$ threshold voltages, respectively.}
\vspace{-2mm}
\label{tab:condition}
\centering
\begin{tabular}{lcccccc}
\hline
& Cut-off && Triode && Saturation \\\hline
\multirow{2}{*}{$MP$} & \multirow{2}{*}{$V_{INV}>V_{DD}+V_{TP}$} && $V_{INV}<V_{DD}+V_{TP}$ && $V_{INV}<V_{DD}+V_{TP}$ \\
&&& $V_{OUT}>V_{INV}-V_{TP}$ && $V_{OUT}<V_{INV}-V_{TP}$\\
&&&&&\\
\multirow{2}{*}{$MN$} & \multirow{2}{*}{$V_{INV}<V_{SS}+V_{TN}$} && $V_{INV}>V_{SS}+V_{TN}$ && $V_{INV}>V_{SS}+V_{TN}$ \\
&&& $V_{OUT}<V_{INV}-V_{TN}$ && $V_{OUT}>V_{INV}-V_{TN}$\\
\hline
\end{tabular}
\end{table*}

\begin{figure}[t]
\centering
\includegraphics[width=3.4in]{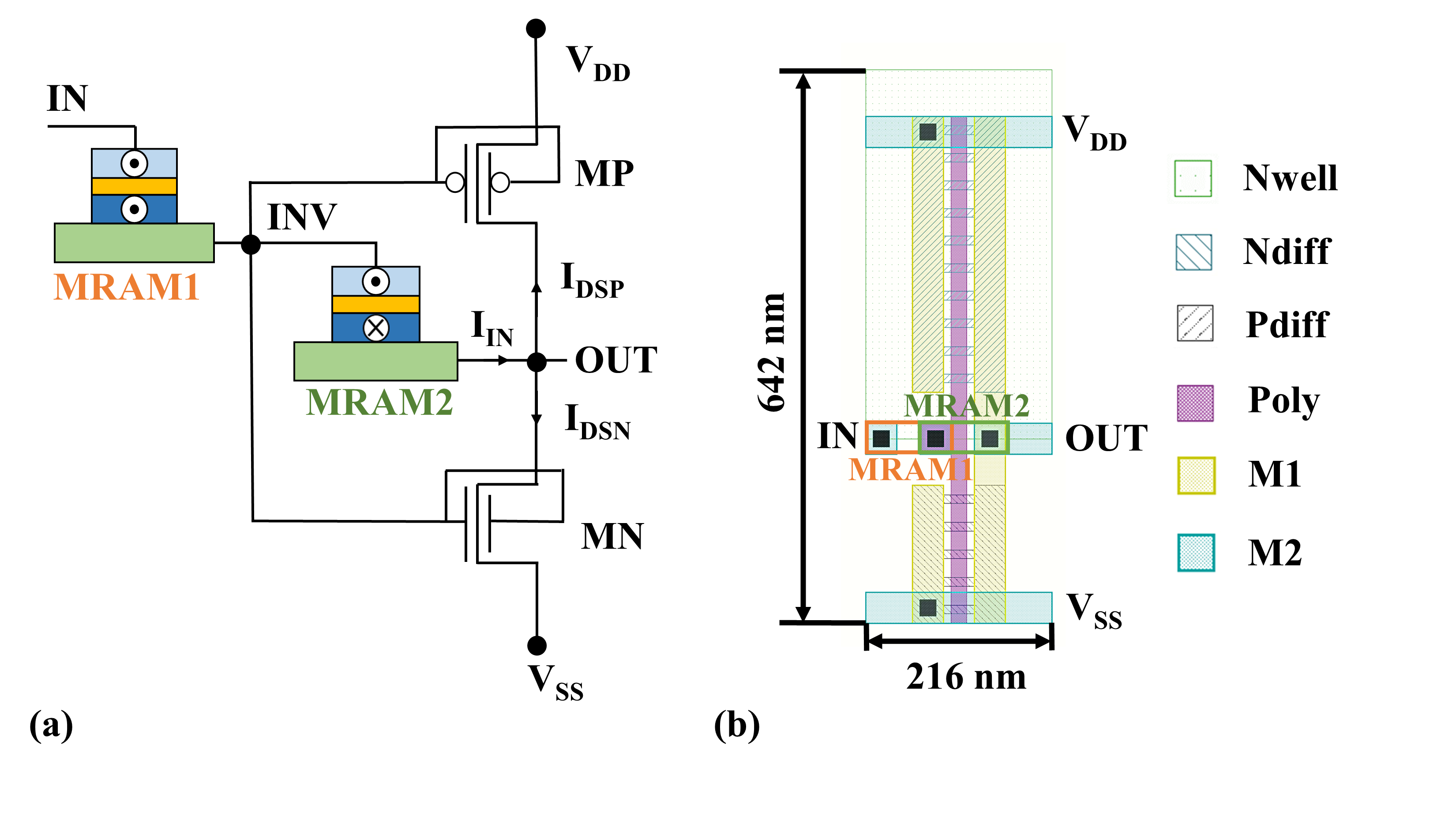}
\caption{(a) Schematic of the proposed analog sigmoid neuron (b) Layout design.}
\label{fig:neuron_sch}
\end{figure}

Our proposed SOT-MRAM-based sigmoid neuron incorporates two SOT-MRAM devices (MRAM1 and MRAM2) and a CMOS-based inverter, as shown in Figure \ref{fig:neuron_sch}. The SOT-MRAM devices are configured in opposite states where MRAM1 is in parallel state ($R_{P}$) and MRAM2 is in anti-parallel state ($R_{AP}$). The SOT-MRAM devices form a voltage divider which reduces the slope of the inverter's linear region leading to a smooth transition from VDD to VSS in the output node (OUT), when the input voltage node (IN) is swept from VSS to VDD. Herein, to verify that such circuit can realize a transcendental activation function between the input voltage ($V_{IN}$) and output voltage ($V_{OUT}$), we conducted a detailed circuit analysis as described in the followings. 

First, we apply the Kirchhoff's current law at OUT node resulting in the following relation:

\begin{equation}
    \label{eqn:main}
    I_{IN}=I_{DSP}+I_{DSN}
\end{equation}
where $I_{IN}$ is the current from node $IN$ to node $OUT$ which passes through the MRAM devices (Fig. \ref{fig:neuron_sch} (a)), and $I_{DSP}$ and $I_{DSN}$ are the drain-to-source currents for the PFET (or PMOS) device $MP$ and NFET (or NMOS) device $MN$, respectively. Considering that no current flows towards the gates of the FinFETs, the $I_{IN}$ can be calculated as follows based on the Ohm's law:
\begin{equation}
    \label{eq:In}
    I_{IN}=\frac{V_{IN}-V_{OUT}}{R_{P}+R_{AP}}
\end{equation}

\noindent where $R_{P}$ and $R_{AP}$ can be obtained from equations (\ref{EqR}) and (\ref{EqTMR}). Replacing $I_{IN}$ in (\ref{eqn:main}) with (\ref{eq:In}), we can calculate $V_{OUT}$ as:
\begin{equation}
    \label{eq:Vout}
    V_{OUT}=V_{IN}-[(R_{P}+R_{AP})(I_{DSP}+I_{DSN})]
\end{equation}

 \begin{figure}[t]
\centering
\includegraphics[width=3.4in]{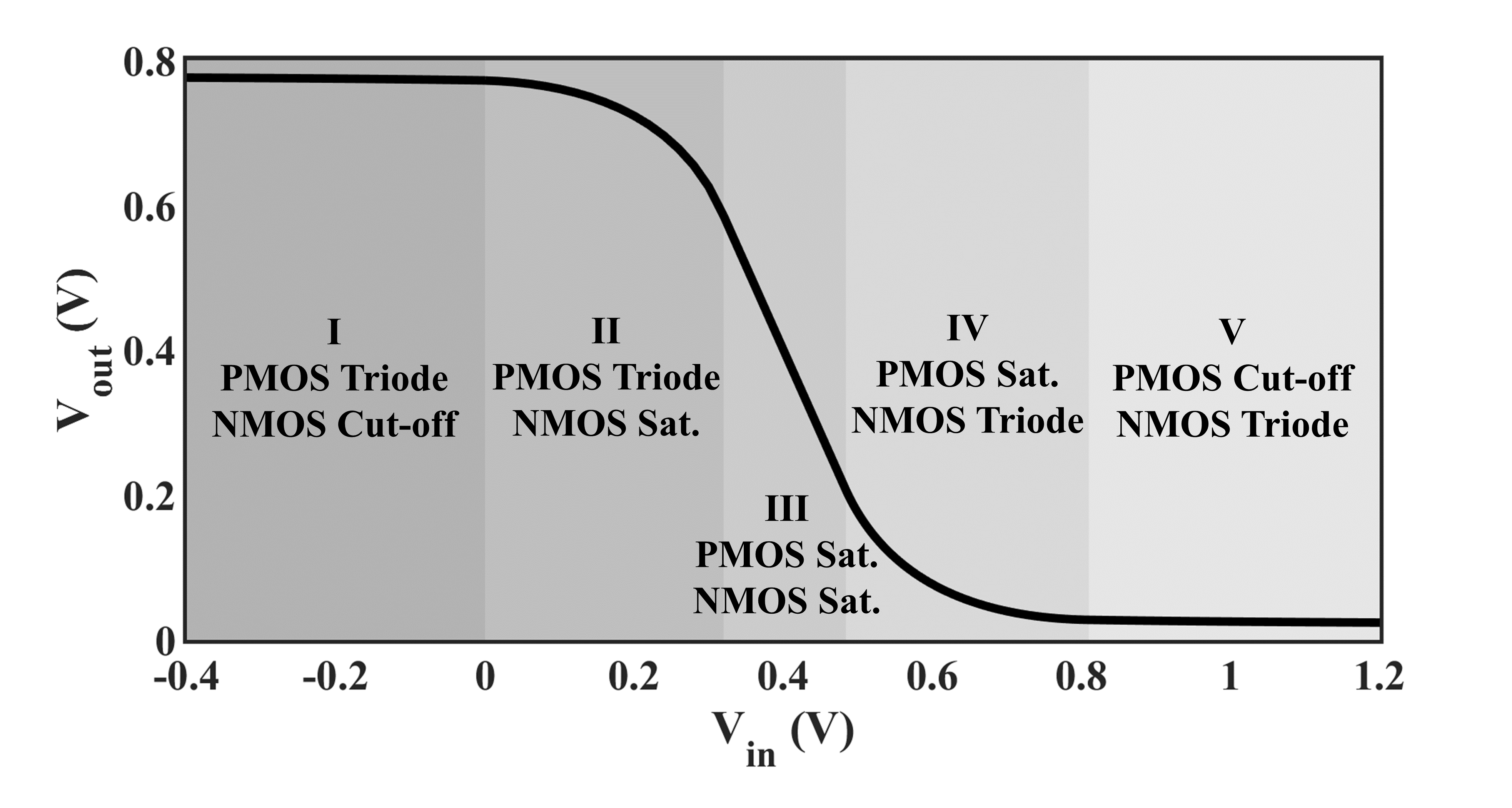}
\caption{VTC of the proposed sigmoid neuron.}
\label{fig:neuron_sig}
\end{figure}

The different regions of operation for our neuron can be identified based on the different operating regions of the $MN$ and $MP$ transistors. To find the operating region of the transistors, we need to calculate their gates' voltage, i.e., $V_{INV}$ in Figure \ref{fig:neuron_sch} (a). Since no current flows towards the gate of the transistors, $V_{INV}$ can be calculated using the below equation:

\begin{equation}
    \label{eq:vinv}
    V_{INV}=V_{IN}-\frac{V_{IN}-V_{OUT}}{R_{P}+R_{AP}}\times R_{P}
\end{equation}

By using equation (\ref{eq:vinv}) and the relations listed in Table \ref{tab:condition}, we can find the operating region of the transistors, and consequently the operating regions of the proposed neuron, based on the input voltage $V_{IN}$. Once the regions are identified, we can use equation (\ref{eq:Vout}) and the $I_{DSP}$ and $I_{DSN}$ relations listed in Table \ref{tab:equation} to find the relation between $V_{OUT}$ and $V_{IN}$. Here, we utilized the MATLAB solver to obtain the relation between the output and input of the neuron. Based on the results, the neuron has five operating regions, the $V_{OUT}$ for each of which is provided in Appendix A for better readability. To exhibit the relation between $V_{OUT}$ and $V_{IN}$, we plotted the voltage transfer curve (VTC) of the neuron based on the equations obtained from the solver (refer to Appendix A) for $VDD$=0.8V and $VSS$=0 as the nominal voltages for the 14nm High-Performance PTM-MG FinFET model \cite{PTM} with $V_{TP}$=-0.2V and $V_{TN}$=0.2V threshold voltages. Figure \ref{fig:neuron_sig} shows the neuron's VTC curve including the five operating regions. As shown, the proposed neuron can provide a transcendental-like activation function that can approximate a $V_{OUT}=sigmoid (-V_{IN})$ function. An actual $V_{OUT}=sigmoid (V_{IN})$ function can also be provided by inverting the $V_{IN}$ before applying it to the proposed neuron. However, as it is described in the next section, inverting the input would not be necessary if the neural network is trained with the inverse sigmoid function.

\section{Results and Discussion}
To investigate the features and benefits of the proposed neuron, we perform a hierarchical simulation process including device-, circuit-, and architecture-level analyses as described in the followings.

\subsection{Device-level Analysis}
\begin{figure}[t]
\centering
\includegraphics[width=3.4in]{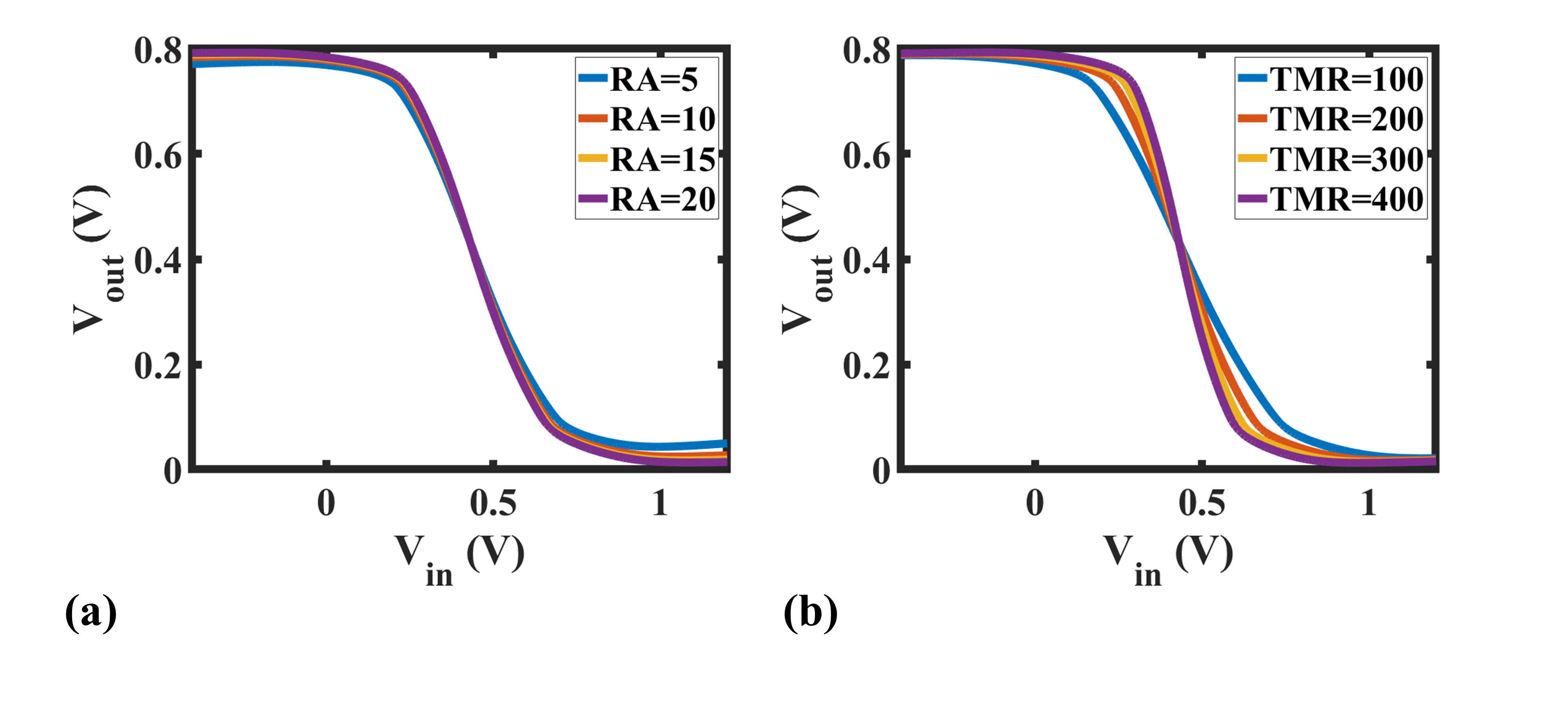}
\vspace{-7mm}
\caption{Effect of varying (a) RA (TMR is fixed to 200) and (b) TMR (RA is fixed to 15 $\Omega \mu m\textsuperscript2$) on the reverse sigmoid output }
\label{fig:tmr_ra}
\end{figure}

To explore the impacts of device-level parameters such as TMR and RA on the shape of the sigmoid function, we implement the proposed neuron in SPICE circuit simulator using the Verilog-A model developed for the SOT-MRAM devices along with 14nm FinFET transistor model. Figure \ref{fig:tmr_ra} shows how changing RA and TMR values in the SOT-MRAM can change the sigmoid function.  

\begin{figure}[t]
\centering
\includegraphics[width=3.4in]{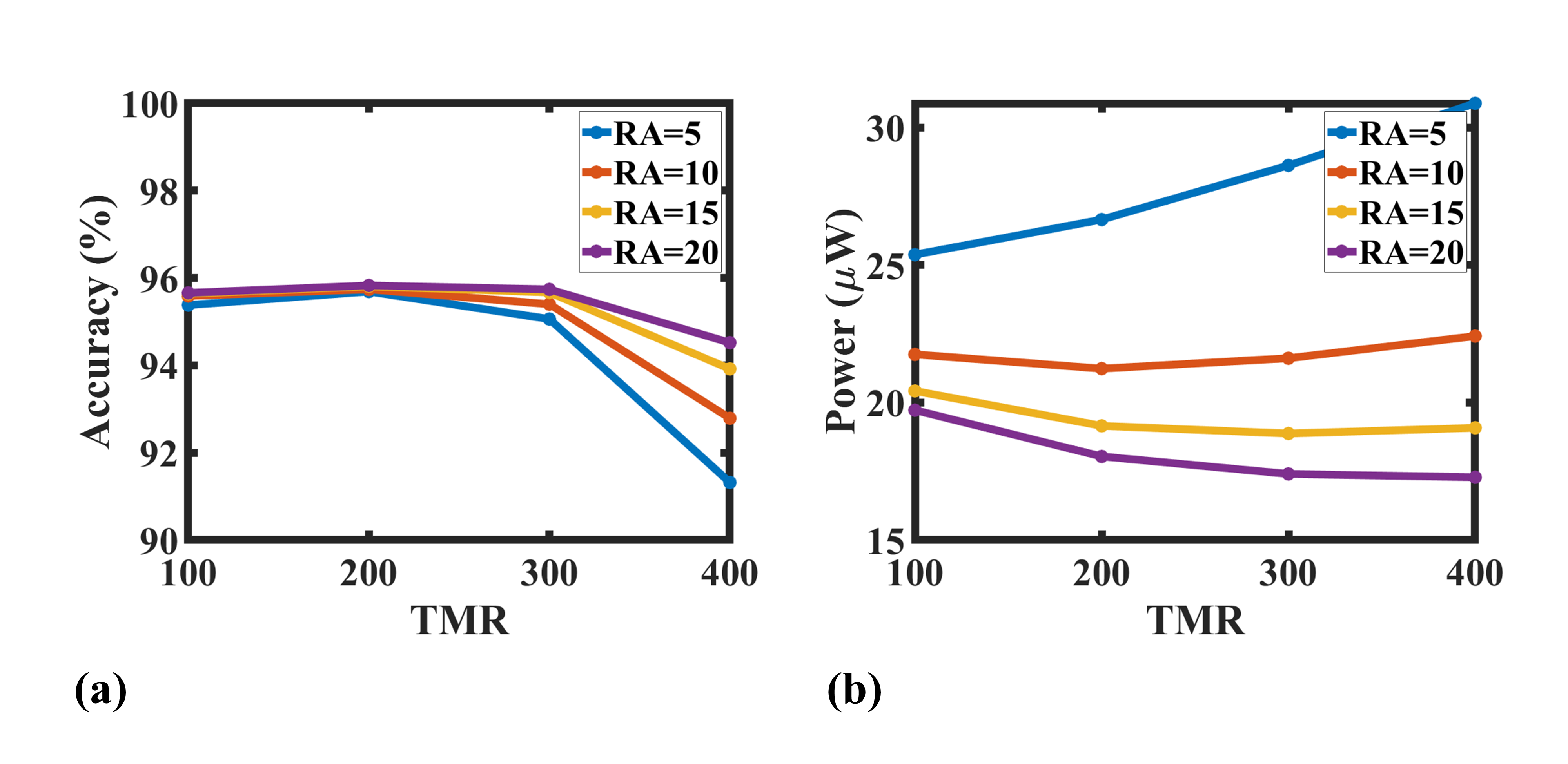}
\caption{(a) MLP classification accuracy, and (b) neuron's power consumption for different values of TMR and RA.}
\label{fig:accuracy}
\end{figure}

To analyze the effect of RA on our neuron, we vary the value of RA to 5, 10, 15 and 20 $\Omega \mu m \textsuperscript2$, while fixing the TMR to 200 (Fig. \ref{fig:tmr_ra} (a)). For analyzing the effect of TMR, we keep the value of RA fixed to 15 $\Omega \mu m\textsuperscript2$ and sweep the value of TMR to 100, 200, 300 and 400, and plot the results in Figure \ref{fig:tmr_ra} (b). The results show that changing TMR imposes larger variations in the shape of the function. However, to investigate how the shape of the sigmoid function can impact the entire network's accuracy, we trained a binarized $400\times120\times84\times10$ MLP model to classify MNIST \cite{MNIST} handwritten images with $20\times20$ pixels. We used $sigmoid(-x)$) activation function during the training to be compatible with the proposed neuron's sigmoidal shape. First, the model is trained for 20 epochs achieving an accuracy of roughly 97\%. Next, the trained weights are mapped to an SOT-MRAM based crossbar. Finally, the proposed SOT-MRAM neuron is integrated with the crossbar and the corresponding IMC circuit is simulated by SPICE. Here, we do not explain the crossbar circuit implementation since it has been already well-investigated in the literature \cite{DPE,parasitic_iscas}. Figure \ref{fig:accuracy} (a) shows how the changes made to the shape of the sigmoid activation function caused by varying the TMR and RA values can impact the overall accuracy of the implemented IMC circuit. The results show that higher RA values consistently lead to better accuracy. However for TMR, the accuracy increases when changing its value from 100 to 200 but it significantly drops when TMR value is increased to 300 and 400. Thus, a maximum accuracy of 95.83\% is achieved with TMR=200 and RA=20 $\Omega \mu m\textsuperscript2$, which is roughly 1\% less than what was realized by ideal software implementation. The change in TMR and RA values can also affect the power consumption of the neuron as shown in Fig. \ref{fig:accuracy} (b), according to which, higher RA and higher TMR values can lead to reduced power consumption.

\subsection{Circuit-level Analysis}

To compare our SOT-MRAM based sigmoid neuron with previous efficient digital \cite{rajput2021vlsi,9218549} and analog \cite{neuron1,neuron2} implementation of transcendental neurons, we conduct SPICE ciruit simulations using 14nm technology node \cite{ptmpaper} and verilog-A model of the SOT-MRAM. For the area comparison, we designed the layout of our neuron as shown in figure \ref{fig:neuron_sch} (b), which demonstrates an area occupation of 0.138 $\mu m^2$. However, as the area occupation values of the previous neurons were measured based on older technology nodes, we use the scaling factors from \cite{Stillmaker2017Scaling7nm} to scale-down their area to 14nm node to provide a fair comparison. For the power consumption comparison, we implemented the previous analog neurons \cite{neuron1,neuron2} and our SOT-MRAM neuron with 14nm technology node. For our neuron, we swept the input voltage from -2V to 2V and measure the average power consumption. However, the previous analog neurons \cite{neuron1,neuron2} take current as input unlike our proposed sigmoid neuron. For fair calculation of the average power consumption, we only consider the active range of their activation function, which we found to be from $-200\mu A$ to $200\mu A$. For the digital neurons, we scaled down their power consumption to 14nm technology using the scaling factors provided in \cite{Stillmaker2017Scaling7nm}. The comparison results are listed in Table \ref{tab:comparison}, according to which our neuron achieves roughly 27$\times$ and 1.8$\times$ power consumption and 4931$\times$ and 2.5$\times$ area reduction compared to the most power- and area-efficient digital and analog neurons, respectively.

\begin{table}[t]
\centering
\caption{Comparison between the SOT-MRAM based neuron and previous digital and analog implementations.}
\small
\label{tab:comparison}
\begin{tabular}{lccccc}
\hline
        & \cite{rajput2021vlsi} & \cite{9218549}& \cite{neuron1} & \cite{neuron2} & This work \\ \hline
Domain  & Digital & Digital & Analog      & Analog     & Analog                                                        \\ \hline
Power ($\mu W$)  & $6.72\times10\textsuperscript{5}$ & 493.4 & 74.21      & 32.16     & 18.04                                                        \\ \hline
Area ($\mu m\textsuperscript{2})$   & 9.3095 & 680.5 & 0.4975       & 0.3505     & 0.138                                                        \\ \hline
\end{tabular}
\end{table}

\subsection{Architecture-level Analysis}

\begin{figure*}[t]
\centering
\includegraphics[width=7in]{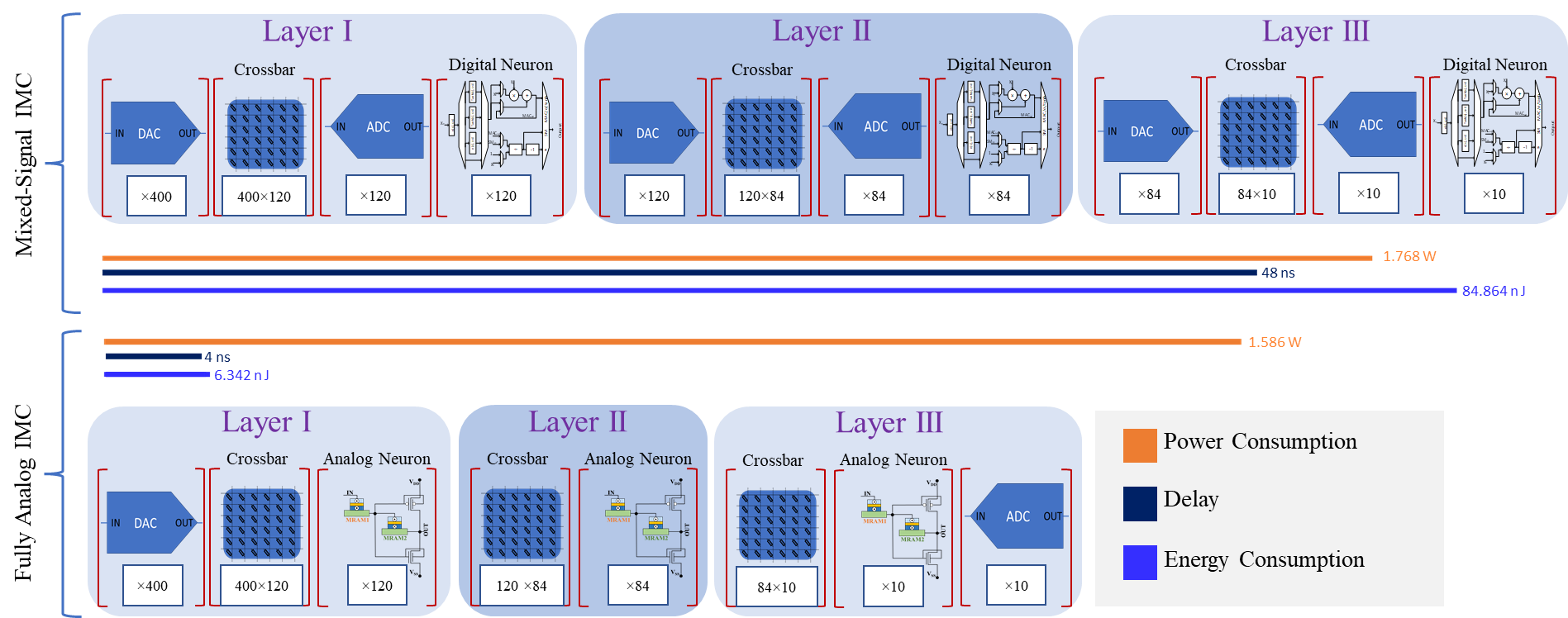}
\caption{The schematic of the mixed-signal and the fully-analog IMC implementations for a $400\times120\times84\times10$ MLP.}
\label{fig:compare}
\end{figure*}

To investigate the benefits of the proposed analog sigmoid at the architecture level, we develop a fully-analog IMC architecture that integrates our SOT-MRAM neuron with an analog memristive crossbar to realize the MVM operations and activations functions in a single fully-connected layer. These analog circuits can be readily integrated to form MLP architectures without requiring intermediate signal conversion units. The only signal converters needed for the fully-analog IMC architecture is a DAC unit in the first layer to convert the digital inputs from a CPU to analog signals, and one ADC unit to convert the analog output of the IMC to digital values to be transferred to the CPU. We compare our design with a conventional mixed-signal IMC architectures which use analog crossbars along with digital neurons, thus imposing the need for DAC and ADC units in every layer. The schematic of both of these designs for implementing the $400\times120\times84\times10$ MLP model discussed in section 4.1 is shown in Figure \ref{fig:compare}, in which the number of neurons, DACs, ADCs and the size of the crossbars in each layer is identified separately.

\begin{table}[]
\caption{Power consumption and latency breakdowns of the fully-analog and mixed-signal implementations of MLP.}
\small
\vspace{-2mm}
\label{tab:sys_comparison}
\centering
\begin{tabular}{cc|cc|cc}
\hline
\multirow{2}{*}{ }                              & \multirow{2}{*}{ } & \multicolumn{2}{c|}{Mixed-Signal} & \multicolumn{2}{c}{Fully-Analog}         \\ \cline{3-6} 
                                                &                   & Power  & Latency                  & Power & Latency \\&       & ($mW$) & ($Clocks$)   & ($mW$) &        ($Clocks$)        \\ \hline
\multicolumn{1}{c|}{\multirow{4}{*}{\rotatebox[origin=c]{90}{Layer 1}}} & DAC       & 3.463                 & \multirow{4}{*}{4} & 3.463                & \multirow{13}{*}{1} \\ \cline{5-5}
\multicolumn{1}{c|}{}                           & Crossbar                          & 221.965                &                   & \multirow{10}{*}{238.405} &                     \\
\multicolumn{1}{c|}{}                           & Neuron                            & 59.208                &                   &                       &                     \\
\multicolumn{1}{c|}{}                           & ADC                               & 5.037                 &                   &                       &                    \\ \cline{1-4}
\multicolumn{1}{c|}{\multirow{4}{*}{\rotatebox[origin=c]{90}{Layer 2}}} & DAC       & 1.039                 & \multirow{4}{*}{4} &                      &                     \\
\multicolumn{1}{c|}{}                           & Crossbar                          & 12.225                &                   &                       &                     \\
\multicolumn{1}{c|}{}                           & Neuron                            & 41.446                &                   &                       &                     \\
\multicolumn{1}{c|}{}                           & ADC                               & 3.526                 &                   &                       &                     \\ \cline{1-4}
\multicolumn{1}{c|}{\multirow{4}{*}{\rotatebox[origin=c]{90}{Layer 3}}} & DAC       & 0.727                 & \multirow{4}{*}{4} &                       &                     \\
\multicolumn{1}{c|}{}                           & Crossbar                          & 1.204                 &                   &                       &                     \\
\multicolumn{1}{c|}{}                           & Neuron                            & 4.934                 &                   &                       &                     \\ \cline{5-5}
\multicolumn{1}{c|}{}                           & ADC                               & 0.42                  &                   & 0.42                     &\\ \hline
\end{tabular}
\end{table}

\begin{table}[]
\caption{Total energy, power, and latency comparison between fully-analog and mixed signal MLP implementations}
\small
\vspace{-2mm}
\label{tab:total_comparison}
\centering
\begin{tabular}{lcccc}
\hline
             & Power($W$) & Latency($ns$) & Energy($nJ$) & TOPS/W \\ \hline
Mixed-Signal & 0.355 & 48 & 17.04 & 3.41 \\
Fully-Analog & 0.242 & 4  &  0.968 & 60.86 \\ \hline
\end{tabular}
\end{table}

Table \ref{tab:sys_comparison} shows the breakdown of the power consumption and latency in the fully-analog IMC and mixed-signal IMC. Here, we utilized a low-power successive-approximation (SAR) ADC and a capacitive DAC proposed in \cite{Kommareddy201910bitCD} for signal conversion. We scaled-down their power consumption from the original 65nm technology to 14nm technology which gives us $41.98\mu W$ and $8.66\mu W$ power consumption for a single ADC and DAC block, respectively. For the mixed-signal implementation, we used the more efficient digital neuron design \cite{neuron2}, which requires three clocks with 250MHz frequency to execute the activation function. Other than the neuron the entire DAC, crossbar, and ADC operations can be completed in one clock cycle. Thus, in the mixed-signal MLP implementation, each layer requires four clock cycles, while the entire fully-analog MLP implementation can be executed in one clock cycles. As for the power consumption, the fully-analog implementation saves power through the elimination of the intermediate conversion units, connecting all crossbars, and the use of power-efficient analog neurons, which are all achieved by using the SOT-MRAM based sigmoid neuron proposed herein. The total power, latency, energy, and performance in terms of tera operations per second per Watt (TOPS/W) of fully-analog and mixed-signal IMC architectures are listed in table \ref{tab:total_comparison}. The results obtained show approximately 13$\times$ improvement in terms of the TOPS/W.

\section{Conclusion}
We proposed a power- and area-efficient SOT-MRAM based neuron circuit realizing an analog sigmoid activation function. Through a hierarchical simulation process, we show the characteristics and benefits of our proposed neuron across device, circuit, and architecture levels. Besides the direct performance and power benefits that can be achieved by the proposed neuron compared to the state-of-the-art, it is shown that it can be readily integrated with analog memristive crossbars without requiring any signal conversion that obviates the need for power-hungry DAC and ADC units. Simulation results exhibited that a fully-analog IMC architecture using our proposed neuron achieves more than $13\times$ improvement in TOPS/W compared to the mixed-signal IMC implementations that utilize analog crossbars along with digital neurons.   

\printbibliography
\appendix

\section{Solving the neuron equations}

Equation \ref{eqn:out1}-\ref{eqn:out5} refers to the calculated analytical solutions for five different operating regions of the proposed sigmoid function, respectively. We define two more intermediate parameters $T_1$ and $T_2$ for better readability of the equations as follows:

\begin{equation}
\begin{split}
    T_1&=\beta_{P}\textsuperscript{2}R_{AP}\textsuperscript{2}(V_{DD}-V_{IN}+V_{TP})\textsuperscript{2}+\beta_{P}\textsuperscript{2}R_{P}\textsuperscript{2}V_{TP}\textsuperscript{2}\\&+2\beta_{P}\textsuperscript{2}R_{AP}R_{P}V_{TP}\textsuperscript{2}+2\beta_{P}R_{P}(V_{DD}-V_{IN}+V_{TP})\\&+2\beta_{P}R_{AP}V_{TP}+1
\end{split}
\end{equation}

\begin{equation}
\begin{split}
    T_2&=\beta_{N}\textsuperscript{2}R_{AP}\textsuperscript{2}(V_{IN}-V_{TN})\textsuperscript{2}+\beta_{N}\textsuperscript{2}R_{P}\textsuperscript{2}V_{TN}\textsuperscript{2}\\&+2\beta_{N}\textsuperscript{2}R_{AP}R_{P}V_{TN}(V_{TN}-V_{IN})-2\beta_{N}R_{AP}V_{TN}\\&+2\beta_{N}R_{P}(V_{IN}-V_{TN})+1
\end{split}
\end{equation}

Equation \ref{eqn:out1} refers to the output of the region I from figure \ref{fig:neuron_sig} where MP and MN stay in the triode and cut-off region, respectively.

\begin{equation}
\label{eqn:out1}
    V_{OUT} =\frac{-1-\beta_{P}R_{P}(V_{DD}+V_{TP})+\beta_{P}R_{AP}(V_{IN}-V_{TP})+T_1\textsuperscript{1/2}}{\beta_{P}(R_{AP}-R_{P})}
\end{equation}

In region II, MN switches to the saturation state and MP remains in the triode state. Equation \ref{eqn:out2} refers to the output for the region II.

\begin{equation}
\label{eqn:out2}
\begin{split}
    V_{OUT}&=[R_{P}-R_{AP}-\beta_{P}R_{P}\textsuperscript{2}(V_{DD}-V_{TN}+V_{TP})\\&-\beta_{P}R_{AP}\textsuperscript{2}(V_{IN}-V_{TP})+\beta_{P}R_{AP}R_{P}(V_{DD}-V_{IN})\\&+\beta_{N}R_{AP}R_{P}(V_{IN}-V_{TN})+(R_{AP}+R_{P})\{T_1\\&-\beta_{N}\beta_{P}R_{AP}\textsuperscript{2}(V_{IN}-V_{TN})\textsuperscript{2}-\beta_{N}\beta_{P}R_{P}\textsuperscript{2}((V_{DD}-V_{TN})\textsuperscript{2}\\&+V_{TP}(V_{DD}-V_{TN}))+2\beta_{N}R_{P}(V_{IN}-V_{TN})\}\textsuperscript{1/2}]\\&/(\beta_{P}R_{AP}\textsuperscript{2}-\beta_{P}R_{P}\textsuperscript{2}+\beta_{N}R_{P}\textsuperscript{2})
\end{split}
\end{equation}

In the next region, MN and MP both work in the saturation region. Equation \ref{eqn:out3} refers to the output equation for the region III.

\begin{equation}
\label{eqn:out3}
\begin{split}
    V_{OUT}&=[R_{P}-R_{AP}-\beta_{P}R_{P}\textsuperscript{2}(V_{DD}-V_{TN}+V_{TP})\\&-\beta_{P}R_{AP}R_{P}(V_{DD}-V_{IN})-\beta_{N}R_{AP}R_{P}(V_{IN}-V_{TN})\\&+(R_{AP}+R_{P})\{2\beta_{P}R_{P}(V_{DD}-V_{IN}+V_{TP})-2\beta_{N}R_{P}V_{TN}\\&-\beta_{N}\beta_{P}R_{P}\textsuperscript{2}((V_{DD}-V_{TN})\textsuperscript{2}+V_{TP}(V_{TP}-2V_{TN}+2V_{DD}))\\&+1\}\textsuperscript{1/2}]/(\beta_{N}R_{P}\textsuperscript{2}-\beta_{P}R_{P}\textsuperscript{2})
\end{split}
\end{equation}

At region IV, MN switches to the triode region while MP stays in the saturation region. (\ref{eqn:out4}) refers to the output for this region:
\begin{equation}
\label{eqn:out4}
\begin{split}
    V_{OUT}&=[R_{P}+R_{AP}-\beta_{P}R_{P}\textsuperscript{2}(V_{DD}+V_{TP})+\beta_{N}R_{P}\textsuperscript{2}V_{TN}\\&-\beta_{N}R_{AP}\textsuperscript{2}(V_{IN}-V_{TN})+\beta_{P}R_{AP}R_{P}(V_{DD}-V_{IN}+V_{TP})\\&+\beta_{N}R_{AP}R_{P}(V_{IN}-2V_{TN})+(R_{AP}+R_{P})\{T_2\\&-\beta_{N}\beta_{P}R_{AP}\textsuperscript{2}(V_{DD}\textsuperscript{2}+V_{TP}\textsuperscript{2}-2V_{DD}V_{IN}+V_{TP}(V_{DD}-V_{IN}))\\&-\beta_{N}\beta_{P}R_{P}\textsuperscript{2}((V_{DD}+V_{TP})\textsuperscript{2}-V_{TN}(V_{DD}+V_{TP}))\\&+2\beta_{P}R_{P}(V_{DD}-V_{IN}+V_{TP})+2\beta_{P}R_{AP}V_{TP}\}\textsuperscript{1/2}]\\&/(\beta_{P}R_{AP}\textsuperscript{2}-\beta_{P}R_{P}\textsuperscript{2}+\beta_{N}R_{P}\textsuperscript{2})
\end{split}
\end{equation}

Finally, MP switches to the cut-off region while MN remains in the triode region. Equation \ref{eqn:out5} refers to the output of the region V.

\begin{equation}
\label{eqn:out5}
    V_{OUT}=\frac{1+\beta_{N}R_{AP}(V_{IN}-V_{TN})-\beta_{N}R_{P}V_{TN}-T_2\textsuperscript{1/2}}{\beta_{N}(R_{AP}-R_{P})}
\end{equation}

\end{document}